\documentclass[aip,onecolumn,showpacs]{revtex4-1}

\usepackage[utf8]{inputenc}
\usepackage{longtable}
\usepackage{subfigure}
\usepackage[tbtags]{amsmath}
\usepackage{booktabs}
\usepackage{makecell}
\usepackage{diagbox}
\usepackage{lastpage}
\usepackage{amssymb,bm,mathrsfs,bbm,amscd}

\usepackage{dcolumn,booktabs}

\newcommand{\TC} {\mbox{\scriptsize TC}}

\newcommand{\vk} {\mathbf{k}}

\def \fuu  {{\left(\mathcal{F}(\triangledown u)^{2}\right)}}

\begin{document}

\title{Perturbation Calculation of the Uniform Electron Gas with a Transcorrelated Hamiltonian}

\author{Hongjun Luo}
\email{h.luo@fkf.mpg.de}
\affiliation{Max-Planck-Institute for Solid State Research, Heisenbergstra\ss e 1, 70569 Stuttgart, Germany}
\author{Ali Alavi}
\email{a.alavi@fkf.mpg.de}
\affiliation{Max-Planck-Institute for Solid State Research, Heisenbergstra\ss e 1, 70569 Stuttgart, Germany}
\affiliation{Department of Chemistry, University of Cambridge, Lensfield Road, Cambridge, CB2 1EW, United Kingdom}


\begin{abstract}
\noindent  
With a transcorrelated Hamiltonian, we perform a many body perturbation (MBPT) calculation on the uniform electron gas in the high density regime.  By using a correlation factor optimised for a single determinant Jastrow ansatz, the second order correlation energy is calculated as 
$\frac{1-\ln2}{\pi^{2}}\ln(r_{s})-0.05075$. This already reproduces the exact logarithmic term of the random phase approximation (RPA) result, while the constant term is roughly $7\%$ larger than the RPA one. The close agreement with the RPA method demonstrates that the transcorrelated  method 
offers a viable and potentially efficient method for treating metallic systems.  

\end{abstract}

\maketitle

\section{Introduction}  
\label{introduction}
Electronic structure calculations usually suffer from numerical problems originating from the Coulomb 
potential ${r_{12}^{-1}}$. On the one hand, the  short-range singularity leads to a non-smoothness in the many body wave function, characterised by a cusp at the electron coalescence point $r_{12}=0$ \cite{Kato57}, which causes a slow convergence with respect to the basis-set in configuration expansions of the wave function . On the other hand, the slow decay of the long-range tail of the Coulomb potential also causes a major problem in calculations of metallic systems.  Any straightforward 
perturbation treatment leads to divergent values in the thermodynamic limit, and in order to get a meaningful perturbation result one has to sum over some sub-sequences of the perturbation expansion up to infinite order, a technique which is usually referred to as the random phase approximation (RPA). \cite{Bohm1951,Pines1952,Bohm1953,Gell-Mann1957,Ehrenreich1959}
 
One way to deal with the short-range singularity of the wave function is to make use of the Jastrow ansatz \cite{jastrow55}
\begin{equation}
    \Psi =e^{\tau} \Phi, \label{jast}
\end{equation}
to split the cusp into a pair correlation factor
\begin{equation}
   \tau(\mathbf{r}_1,\cdots,\mathbf{r}_N)=
   \frac 1 2\sum_{i,j} u(\mathbf{r}_i,\mathbf{r}_j).
\end{equation}
With a factor $u$ fulfilling the asymptotic condition
\begin{equation}
    u(\mathbf{r}_i,\mathbf{r}_j)\sim \frac 1 2 |\mathbf{r}_i-\mathbf{r}_j|,\quad, \mbox{when} 
  \   |\mathbf{r}_i-\mathbf{r}_j|\rightarrow 0,
\end{equation}
the smoothness of the function $\Phi$ is one order higher than that of the function $\Psi$ \cite{FHHO05}. This product form is widely used in quantum Monte Carlo methods  \cite{ceperley78,umrigar93,foulkes01}, where the expectation value of the Hamiltonian can be evaluated by random sampling. One can thus minimise this expectation value of energy by adjusting the correlation factor $\tau$. 
Due to the involved high dimensional integrals, it is difficult to use such kind of variational treatment of the Jastrow wave function in conventional quantum chemistry, where the function $\Phi$ needs to be further approximated by configuration expansions. Alternatively, the transcorrelated (TC) method of Boys and Handy \cite{Boys69, 
Boys69a, Boys69b, Boys69c,  Boys69d, Handy69, Handy72, Handy73} offers a fairly efficient way to deal with the Jastrow ansatz.  With this method, the original Schr\"odinger equation is transformed into a non-Hermitian eigenvalue problem 
\begin{equation}
    \hat{H}_{\TC}\Phi=E\Phi, 
    \qquad \hat{H}_{\TC}=e^{-\tau}\hat{H}e^{\tau},
\end{equation}
via an exact similarity transformation, which removes the involved exponential factor, and results in an exact effective Hamiltonian containing up to three body terms: 
\begin{eqnarray}
    \hat{H}_{\TC}&=& \hat{H}+[\hat{H},\tau]+\frac 1 2 [[\hat{H},\tau],\tau] \nonumber\\
    &=&\hat{H}- \sum_i \left( \frac 1 2 \triangledown_i ^2 \tau +(\triangledown_i \tau)\cdot \triangledown_i +\frac 1 2 (\triangledown_i \tau)^2  \right).
\label{eq_HTC}
\end{eqnarray}
This method was initially studied by Boys and Handy for the single determinant ansatz, and has only recently been combined with configuration expansion methods of quantum chemistry\cite{Ten-no00,Hino01,Hino02,Luo11,Luo2018,Cohen2019,Guther2021,Liao2021,Schraivogel2021}. These works reveal a high level of efficiency of the transcorrelated Hamiltonian in handling short range correlations, that the resolving of the cusp leads to significant speed up of the basis convergence, and further, more speed up can be achieved when more generalised factors are used (instead of the simple $r_{12}$ type or $F(r_{12})$ type factors).

Regarding the study of long-range correlation with the transcorrelated method, there exists only a few works dealing with the single determinant ansatz for uniform electron gases (UEG) \cite{Armour80,Umezawa04,luo12}. By choosing a proper correlation factor $\tau$, the leading order singularity of the Coulomb potential can be removed from the effective Hamiltonian, both for the short-range and the long-range singularities. We therefore expect  
that the divergence of the perturbation energy for metallic systems can be cured to some extent. At least it should be possible to get non-divergent results in the thermodynamic limit at low order perturbation theory without any RPA-type treatment. It would then be interesting to know whether this kind of non-RPA treatment leads to meaningful results. In this work, we will test the performance of a MBPT treatment of the transcorrelated Hamiltonian on UEG in the high density regime, which has a well established RPA result \cite{Gell-Mann1957,Onsager1966}
\begin{equation}
    \frac {E_c} N=\frac{1-\ln 2}{\pi ^2}\ln{r_s} -0.047 +\cdots.
\end{equation}

In the next section, we describe the perturbation treatment of the TC Hamiltonian, and in sections \ref{1st} and \ref{2nd} we present details of the calculations of the first and the second order energies, where a Jastrow factor optimised for the Hartree-Fock reference function is used.
In section \ref{CY}, we also present some calculation results with the widely used Coulomb-Yukawa factor,  which will be followed by some concluding remarks in section \ref{conclud}.

\section{Perurbation treatment of the transcorrelated Hamiltonian}
Uniform electron gases are described by the Hamiltonian
\begin{equation}
    \hat{{\cal H}}=-\frac{1}{2}\sum_{i}\triangledown_{i}^{2}+\sum_{i<j}\frac{1}{r_{ij}}.
\end{equation}
The system is characterized by the charge density $\rho$ (or equivalently the Wigner–Seitz radius $r_{s}$,
or the Fermi wavevector $k_{F}$). In order to get the explicit dependence of the physical quantities
on $r_{s}$, we re-scale the spatial coordinates 
\begin{equation}
    {\bf r}_{i}=\alpha r_{s}{\bf r}'_{i},\quad 
    \alpha=\left(\frac{4}{9\pi}\right)^{\frac{1}{3}},
\end{equation}
so that the system keeps a fixed density
\begin{equation}
    r_{s}^{\prime}=\frac{1}{\alpha},\ k_{F}^{\prime}=1,\rho^{\prime}=\frac N \Omega =\frac{1}{3\pi^{2}},
    \label{eq_rs_kf}
\end{equation}
but has a re-scaled Hamiltonian
\begin{eqnarray}
\hat{{\cal H}} & = & \frac{1}{\alpha^{2}r_{s}^{2}}\hat{H},\label{eq_rescal}\\
\hat{H} & = & -\frac{1}{2}\sum_{i}\triangledown_{i}^{2}+\alpha r_{s}\sum_{i<j}\frac{1}{r_{ij}}. \label{eq_rescale_2}
\end{eqnarray}
The formal volume $\Omega$ and the particle number $N$ will finally be taken as $\infty$ in the thermodynamic limit.
In this paper we will only use this re-scaled form and thus will ignore the prime on the variables. 
We need only to remember that the final energies should be divided by $\alpha ^2 r_s ^2$ according to equation (\ref{eq_rescal}). 
Then we use Jastrow ansatz (\ref{jast}) for the ground state wave function of $\hat{H}$, and the two-body correlation factor $u(\mathbf{r}_i,\mathbf{r}_j)$ can be written as  $u(\mathbf{r}_i-\mathbf{r}_j)$ due to the translational symmetry. 
The transcorrelated Hamiltonian (\ref{eq_HTC}) can be derived straightforwardly in the second quantisation formalism \cite{luo12} in terms of plane wave orbitals
\begin{eqnarray}
\hat{H}_{\TC}&=&e^{-\tau}\hat{H}e^{\tau}=\hat{H}+\hat{K}+\hat{L},\\
\hat{K} & = & \frac{1}{2\Omega}\sum_{\sigma\sigma'}\sum_{kpq}\left(k^{2}\tilde{u}(\mathbf{k})-(\mathbf{p}-\mathbf{q})\cdot\mathbf{k}\ \tilde{u}(\mathbf{k}) 
 -\fuu (\mathbf{k})\right) \nonumber\\
 && a_{\mathbf{p}-\vk,\sigma}^{\dagger}a_{\mathbf{q}+\mathbf{k},\sigma'}^{\dagger}a_{\mathbf{q},\sigma'}a_{\mathbf{p},\sigma}, \label{eq_K}\\
\hat{L} & = & -\frac{1}{2\Omega^{2}}\sum_{\sigma\sigma'\sigma''}\sum_{kk'pqs}\tilde{u}(\mathbf{k})\tilde{u}(\mathbf{k}')\mathbf{k}'\cdot\mathbf{k}\ a_{\mathbf{p}-\mathbf{k},\sigma}^{\dagger}a_{\mathbf{q}+\mathbf{k}',\sigma'}^{\dagger}a_{\mathbf{s}+\mathbf{k}-\mathbf{k}',\sigma''}^{\dagger}a_{\mathbf{s},\sigma''}a_{\mathbf{q},\sigma'}a_{\mathbf{p},\sigma},\label{eq_L}
\end{eqnarray}
where $\tilde{u}(\mathbf{k})$ is the Fourier transformation of $u(\mathbf{r})$.  $\fuu (\mathbf{k})$ is the Fourier transformation of $(\triangledown u(\mathbf{r}))^2$ and can be calculated via the convolution theorem
\begin{equation}
   \fuu (\mathbf{k})=-\frac{1}{(2\pi)^3}\int (\vk -\vk')\cdot \vk'
    \tilde{u}(|\vk-\vk'|)\tilde{u}(k')d^3k'.
\end{equation}
Here we have used the fact that in the thermodynamic limit, the function $\tilde{u}(\vk)$ is actually a 1D function of $k$ due to rotational symmetry.  

It is of crucial importance to take a proper Jastrow factor $u$.  It has to fulfil asymptotic conditions at both of the limits \cite{luo12}
\begin{eqnarray}
\tilde{u}(k) &\sim& -\frac {w(k)}{k^2} \qquad \mbox{when } k\sim \infty ,\\
\tilde{u}(k) &\sim & -\frac{\sqrt{w(k)}}{\sqrt{\rho}k} \qquad \mbox{when } k\sim 0, 
\end{eqnarray}
where $w(k) = \frac{4\pi \alpha r_s}{k^2}$ is the Fourier transformation of the Coulomb potential and $\rho$ is the constant density given in equation (\ref{eq_rs_kf}). These conditions remove the leading order singular term of the Coulomb potential and lead to a finite second order perturbation energy. However, they do not guarantee the {\em quality} of the perturbation energy. Based on some recent numerical studies of the TC method on small molecular systems \cite{Cohen2019,Guther2021}, the correlation factor optimised for a single determinant reference $\Phi$ (e.g., the Hartree-Fock wave function) also performs well when $\Phi$ is further treated by configuration expansions, at least for weakly correlated systems. For UEG in the high density limit, the correlation factor optimised for the Hartree-Fock reference takes the following form
\begin{equation}
    \tilde{u}(k)=\frac{k^2-\sqrt{k^4+4\rho w(k) k^2 T^2_2(k)}}{2\rho k^2T_2(k)}.
    \label{eq_u}
\end{equation}
$T_2(k)$ is a volume factor defined by
\begin{equation}
    T_2(k)\equiv \frac 2 N \sum_\mathbf{p} \Theta(1-p)\Theta(|\mathbf{p}-\vk|-1),
\end{equation}
with $\Theta$ being the Heaviside step function. In the thermodynamic limit, $T_2(k)$ can be calculated as
\begin{equation}
    T_2(k)=\begin{cases}
            1, & k>2,\\
            \frac 3 4 k-\frac 1 {16} k^3, & k\le 2.
            \end{cases}
\end{equation}
The expression (\ref{eq_u}) of $\tilde{u}(k)$ is a solution to a quadratic equation
\begin{equation}
    \rho k^2 T_2(k)\tilde{u}^2(k)-k^2\tilde{u}(k)-w(k)T_2(k)=0, \label{eq_quadratic}
\end{equation}
which can can be derived both by a variational treatment of the Jastrow ansatz \cite{Gaskell61,Gaskell62,Talman74} and by transcorrelated method \cite{Armour80,luo12}.

The Hamiltonian (\ref{eq_HTC}) is then split into two parts for a perturbation treatment
\begin{equation}
    \hat{H}_{\TC}=\hat{H}_0+\hat{W}_1,\qquad \hat{W}_1=\hat{W}+\hat{K}+\hat{L},
    \label{eq_H_pert}
\end{equation}
where for simplicity we take only the kinetic terms in $\hat{H}_0$
\begin{eqnarray}
\hat{H}_0&=&\sum_{p\sigma} \frac 1 2 p^2 \hat{n}_{\mathbf{p},\sigma},\\
\hat{W}&=& \frac{1}{2\Omega}\sum_{\sigma\sigma'}\sum_{kpq}\ w(k)\ a_{\mathbf{p}-\mathbf{k},\sigma}^{\dagger}a_{\mathbf{q}+\mathbf{k},\sigma'}^{\dagger}a_{\mathbf{q},\sigma'}a_{\mathbf{p},\sigma}.
\end{eqnarray}
With this partition of the TC Hamiltonian, the perturbation energies can be written as
\begin{eqnarray}
E_{0} & = & \langle\Phi_{0}|\hat{H}_{0}|\Phi_{0}\rangle,\\
E_{1} & = & \langle\Phi_{0}|\hat{W}_{1}|\Phi_{0}\rangle,\\
E_{2} & = & -\langle\Phi_{0}|(\hat{W}_{1}-E_{1})\frac{1}{\hat{H}_{0}-E_{0}}(\hat{W}_{1}-E_{1})|\Phi_{0}\rangle,
\label{eq_E2}
\end{eqnarray}
where $E_0$ is simply the total kinetic energy and $E_1$ contains the exchange energy $\langle\Phi_{0}|\hat{W}|\Phi_{0}\rangle$ and the first order correlation energy $\langle\Phi_{0}|\hat{K}+\hat{L}|\Phi_{0}\rangle$. The second order energy contributes only to the correlation energy.

\section{The first order perturbation energy}\label{1st}
The first order correlation energy is composed of the expectation values of the two body operator $\hat{K}$ and the three body operator $\hat{L}$. For calculations of these values, we need to find all possible contractions of the creation and annihilation operators. From the two body operator we get a direct  and an exchange contribution
\begin{equation}
   \langle\Phi|\hat{K}|\Phi\rangle= -\frac{N^{2}}{2\Omega}\fuu (\mathbf{0})
    +\frac{N}{2\Omega} \sum_{\vk} \fuu (\vk)\left(1-T_{2}(k)\right).
    \label{eq_E1_K}
\end{equation}
Here only the $\fuu$ term of $\hat{K}$ makes a non-vanishing contribution, while the other terms coming from 
$[\hat{H},\tau]$ have no contribution to the expectation value. The  expectation value of the three body operator also has two possible contractions
\begin{equation}
    <\Phi|\hat{L}|\Phi>=\frac{N^{2}}{2\Omega^{2}}\sum_{\mathbf{k}}\left(1-T_{2}(k)\right)k^{2}\tilde{u}^{2}(\mathbf{k})-\frac{N}{\Omega^{2}}\sum_{\mathbf{k}\mathbf{k}'}\tilde{u}(\mathbf{k})\tilde{u}(\mathbf{k}')\mathbf{k}\cdot\mathbf{k}'O_{3}(\mathbf{k},\mathbf{k}').
    \label{eq_E1_L}
\end{equation}
The volume factor $O_3$ is defined as
\begin{equation}
    O_3(\vk,\vk')=\frac 2 N \sum _\mathbf{p} \Theta(1-p)\Theta(1-|\mathbf{p}-\vk|)\Theta(1-|\mathbf{p}-\vk'|),
\end{equation}
which can be calculated analytically in the thermodynamic limit\cite{luo12}.  

In the thermodynamic limit the summation should be replaced by a integration 
\begin{equation}
\frac 1 \Omega \sum_\vk\rightarrow \frac 1 {(2\pi)^3} \int d^3k.
\end{equation}
This replacement will be used in the following context without any notification.

The first term in equation (\ref{eq_E1_L}) is partly cancelled by the first term in equation (\ref{eq_E1_K}) and we have finally three terms in the expression of the first order correlation energy
\begin{eqnarray}
E_{1}^{c} & = & E_{11}^{c}+E_{12}^{c}+E_{13}^{c},\\
E_{11}^{c} & = & -\frac{\rho N}{2(2\pi)^{3}}\int T_{2}(k)\mathbf{k}^{2}\tilde{u}^{2}(\mathbf{k})d^{3}k,
\label{eq_E11_0} \\
E_{12}^{c} & = & \frac{N}{2(2\pi)^{3}}\int d^{3}k\left({\cal F}((\triangledown u)^{2})(\mathbf{k})\left(1-T_{2}(k)\right)\right),\\
E_{13}^{c} & = & -\frac{N}{(2\pi)^{6}}\int d^{3}kd^{3}k'\ \tilde{u}(\mathbf{k})\tilde{u}(\mathbf{k}')\mathbf{k}\cdot\mathbf{k}'O_{3}(\mathbf{k},\mathbf{k}').
\end{eqnarray}
For a more transparent treatment, we reformulate the expression (\ref{eq_u}) of $\tilde{u}(k)$ as
\begin{equation}
    \tilde{u}(k) = -\alpha r_{s}\frac{8\pi T_{2}(k)}{k^{2}\left(k^{2}+\sqrt{k^{4}+\frac{16}{3\pi}\alpha r_{s}T_{2}^{2}(k)}\right)}.
    \label{eq_u_2}
\end{equation}
The expression of the first order energy contains only quadratic terms of $\tilde{u}(k)$ and the prefactor $\alpha r_s$ of $\tilde{u}$ in equation (\ref{eq_u_2}) will be cancelled by the prefactor $\frac{1}{(\alpha r_s)^2}$ in the final energy expression according to equation (\ref{eq_rescal}). If $r_s$ is set to be $0$ in the remaining integrands in the first order energy, it is found that $E^c_{11}$ will diverge while the expression of $E^c_{12}$ and $E^c_{13}$ still give finite results. 
The singular term in $E^c_{11}$ originates from the integration over small $k$. In order to get a analytic result of the singular term we rearrange the expression of $E_{11}^c$ as
\begin{eqnarray}
\frac{E_{11}^c} {N} & = & -\frac{1}{6\pi^{2}(2\pi)^{3}}\int T_{2}(k)k^{2}\tilde{u}^{2}(k)d^{3}k\\
 & = & -\frac{3}{16}\int_{0}^{2}T_{2}(k)\left(\frac{k^{2}-\sqrt{k^{4}+\frac{16}{3\pi}\alpha r_{s}T_{2}^{2}(k)}}{T_{2}(k)}\right)^{2}dk-\frac{4\pi}{6\pi^{2}(2\pi)^{3}}\int_{2}^{\infty}T_{2}(k)k^{4}\tilde{u}^{2}(k)dk \nonumber \\
 & = & -\frac{3}{16}\int_{0}^{2}\frac{3k}{4}\left(\frac{k^{2}-\sqrt{k^{4}+\frac{16}{3\pi}\alpha r_{s}(\frac{3}{4}k)^{2}}}{\frac{3}{4}k}\right)^{2}dk \nonumber\\
 &  & -\frac{3}{16}\int_{0}^{2}\left[T_{2}(k)\left(\frac {\frac{16}{3\pi}\alpha r_{s}T_{2}(k)} {k^{2}+\sqrt{k^{4}+\frac{16}{3\pi}\alpha r_{s}T_{2}^{2}(k)}} \right)^{2}-\frac{3k}{4}
 \left(\frac {\frac{16}{3\pi}\alpha r_s (\frac{3}{4}k)} {k^{2}+\sqrt{k^{4}+\frac{16}{3\pi}\alpha r_{s}(\frac{3}{4}k)^2}} \right)^{2}\right]dk \nonumber\\
 &  & -\frac{1}{12\pi^{4}}\int_{2}^{\infty}T_{2}(k)k^{4}\tilde{u}^{2}(k)dk.
 \label{eq_E11}
\end{eqnarray}
In the first line of expression (\ref{eq_E11}) $T_2(k)$ is simplified with its leading term $\frac 3 4 k$ in the small k region, and this simplified integrand  is subtracted out from the original integrand in the second line. The simplified integral in the first line can be calculated analytically as
\begin{eqnarray}
&&-\frac{3}{16}\int_{0}^{2}\frac{3k}{4}\left(\frac{k^{2}-\sqrt{k^{4}+\frac{16}{3\pi}\alpha r_{s}(\frac{3}{4}k)^{2}}}{\frac{3}{4}k}\right)^{2}dk\nonumber\\
&=& \alpha^{2}r_{s}^{2}\left[\frac{9}{32\pi^{2}}\ln(r_{s})+\frac{1}{16}\left(\frac{3}{\pi}\right)^{2}\left(\ln\left(2\sqrt{\frac{3\alpha}{\pi}}\right)-\ln8\right)+\frac{1}{64}\left(\frac{3}{\pi}\right)^{2}\right]+o(r_{s}^{2}).
\end{eqnarray}
In the second line, the singularities of the two terms cancel each other and hence they contribute at most only to the constant term in the correlation energy. For the calculation of the contribution to the constant, we can simply take out the common $\alpha^2 r_s^2$ factor and set $r_s=0$ in the remaining integrand, which gives
\begin{eqnarray}
&&-\frac{3}{16}\int_{0}^{2}\left[T_{2}(k)\left(\frac {\frac{16}{3\pi}\alpha r_{s}T_{2}(k)} {k^{2}+\sqrt{k^{4}+\frac{16}{3\pi}\alpha r_{s}T_{2}^{2}(k)}} \right)^{2}-\frac{3k}{4}
 \left(\frac {\frac{16}{3\pi}\alpha r_s (\frac{3}{4}k)} {k^{2}+\sqrt{k^{4}+\frac{16}{3\pi}\alpha r_{s}(\frac{3}{4}k)^2}} \right)^{2}\right]dk \nonumber\\
 &=&  -\alpha^{2}r_{s}^{2}\left(-\frac{8\cdot3^{2}}{16^{2}\pi^{2}}+\frac{12}{16^{2}\pi^{2}}-\frac{1}{9\cdot2^{5}\pi^{2}}\right)+o(r_{s}^{2}).
\end{eqnarray}
In a similar way, the integral in the third line can be easily calculated as
\begin{equation}
   -\frac{1}{12\pi^{4}}\int_{2}^{\infty}T_{2}(k)k^{4}\tilde{u}^{2}(k)dk
   =-\frac{\alpha^{2}r_{s}^{2}}{18\pi^{2}}+o(r_{s}^{2}).
\end{equation}
$E_{12}^c$ and $E_{13}^c$ contribute also only at most to the constant term of the correlation energy and can also be simplified with the above method. After this simplification, it is still too difficult to be calculated analytically due to the complicated integrand. There is, however, no problem for a numerical calculation of such 1D integrals, which leads to
\begin{eqnarray}
\frac{E_{12}^{c}}{N} & = & 0.02352\alpha^{2}r_{s}^{2}+o(r_{s}^{2}),\\
\frac{E_{13}^{c}}{N} & = & -0.01310\alpha^{2}r_{s}^{2}+o(r_{s}^{2}).
\end{eqnarray}
By adding all the results together, we get the final result for the first order correlation energy
\begin{equation}
    \frac{E_{1}^{c}}{N}=\alpha^{2}r_{s}^{2}\left[\frac 9 {32\pi^2}\ln(r_{s})-0.05576\right]+o(r_{s}^{2}).
\end{equation}

The first order correlation energy is essentially the correlation energy produced by the single determinant Jastrow ansatz. It is interesting to find that the leading logarithmic term $\frac 9 {32\pi^2}\ln(r_{s})$ is the same as that obtained by Talman based on linked cluster expansion \cite{Talman74}.   This logarithmic term makes up roughly $90\%$ of the exact RPA logarithmic term $\frac{1-\ln{2}}{\pi^2} \ln r_s$, which is also a typical portion of correlated  energy one may expect from variational quantum Monte Carlo calculations based on single determinant Jastrow ansatz.

\section{The second order perturbation energy}\label{2nd}
The transcorrelated effective potential $\hat{W}_1$ depends on $r_s$ nonlinearly. Besides the original Coulomb potential $w(k)$, which is proportional to $ r_s$, there are some terms which rely on $\tilde{u}(k)$ linearly and others quadratically.  According to equation (\ref{eq_u_2}), in the high density regime where $r_s\sim 0$, $\tilde{u}(k)$ scales roughly linearly with $r_s$. To get an estimation of the contribution to the second order energy $E_2$ from each quadratic term in the effective potential $\hat{W}_1$, we can first set $r_s=0$ in the denominator of the expression (\ref{eq_u_2}) of $\tilde{u}(k)$
\begin{equation}
    \tilde{u}(k)\approx -\frac{4\pi \alpha r_s T_2 (k)}{k^4}. \label{eq_u_simple}
\end{equation} 
If this leads only to finite integrals in the calculation of  $E_2$, we can ignore this quadratic term for the current calculation, since it gives at most a $O(r_s^3)$ contribution to $E_2$ according to equation (\ref{eq_E2}). Based on this estimation we find there is only one quadratic term left, where the approximation (\ref{eq_u_simple}) leads to divergent integrals in the calculation of $E_2$.
This term looks like
\begin{equation}
    -\frac{\rho}{2\Omega}\sum_{\sigma\sigma'}\sum_{kpq}k^2 \tilde{u}^2(\mathbf{k}) a_{\mathbf{p}-\mathbf{k},\sigma}^{\dagger}a_{\mathbf{q}+\mathbf{k},\sigma'}^{\dagger}a_{\mathbf{q},\sigma'}a_{\mathbf{p},\sigma},
\end{equation}
which is a contraction of the three body potential $\hat{L}$ in equation (\ref{eq_L}) by taking $\vk=\vk'$.
Consequently for the calculation of the two leading terms in the second order energy $E_{2}$, we need
only to take the following approximation of of the effective potential
\begin{eqnarray}
\hat{W}_{1} & \approx & \frac{1}{2\Omega}\sum_{\sigma\sigma'}\sum_{kpq}\ G(\mathbf{p},\mathbf{q},\mathbf{k})\ a_{\mathbf{p}-\mathbf{k},\sigma}^{\dagger}a_{\mathbf{q}+\mathbf{k},\sigma'}^{\dagger}a_{\mathbf{q},\sigma'}a_{\mathbf{p},\sigma},\\
G(\mathbf{p},\mathbf{q},\mathbf{k}) & = & G_{0}(k)-(\mathbf{p}-\mathbf{q})\cdot\mathbf{k}\tilde{u}(k), \label{eq:G}\\
G_{0}(k) & = & w(k)+k^{2}\tilde{u}(k)-\rho k^{2}\tilde{u}^{2}(k).\label{eq_G0}
\end{eqnarray}
By taking this two body potential approximation of $\hat{W}_{1}$,
the second order energy consists two contributions
\begin{eqnarray}
E_{2} & = & -\langle\Phi_{0}|(\hat{W}_{1}-E_{1})\frac{1}{\hat{H}_{0}-E_{0}}(\hat{W}_{1}-E_{1})|\Phi_{0}\rangle
\nonumber \\
 & \approx & E_{21}+E_{22},
\end{eqnarray}
obtained by a direct contraction and an exchange  contraction respectively.

The direct contraction term of $E_2$ can be written as
\begin{eqnarray}
E_{21} & = & -\frac{2}{4\Omega^{2}}\sum_{\sigma\sigma'}\sum_{kpq}\frac{G(\mathbf{p},\mathbf{q},\mathbf{k})G(\mathbf{p}-\mathbf{k},\mathbf{q}+\mathbf{k},-\mathbf{k})}{\frac{1}{2}\left((\mathbf{p}-\mathbf{k})^{2}+(\mathbf{q}+\mathbf{k})^{2}-p^{2}-q^{2}\right)}\Theta(1-p)\Theta(1-q)\Theta(|\mathbf{p}-\mathbf{k}|-1)\Theta(|\mathbf{q}+\mathbf{k}|-1)\nonumber \\
 & = & -\frac{4}{\Omega^{2}}\sum_{kpq}\int_{0}^{\infty}dt\left\{ G(\mathbf{p},\mathbf{q},\mathbf{k})G(\mathbf{p}-\mathbf{k},\mathbf{q}+\mathbf{k},-\mathbf{k})\right.\nonumber\\
 &&\qquad \left. \Theta(1-p)\Theta(1-q)\Theta(|\mathbf{p}-\mathbf{k}|-1)\Theta(|\mathbf{q}+\mathbf{k}|-1)e^{-\left(2k^{2}-2(\mathbf{p}-\mathbf{q})\cdot\mathbf{k}\right)t}\right\} \nonumber \\
 & = & -\frac{4}{\Omega^{2}}\sum_{kpq}\int_{0}^{\infty}dt\left\{ \left(\left(G_{0}(k)-k^{2}\tilde{u}(k)\right)^{2}-\left(k^{2}-(\mathbf{p}-\mathbf{q})\cdot\mathbf{k}\right)^{2}\tilde{u}^{2}(k)\right)\right.\nonumber\\
 && \qquad \left.\Theta(1-p)\Theta(1-q)\Theta(|\mathbf{p}-\mathbf{k}|-1)\Theta(|\mathbf{q}+\mathbf{k}|-1)e^{-\left(2k^{2}-2(\mathbf{p}-\mathbf{q})\cdot\mathbf{k}\right)t}\right\} \nonumber \\
 & = & -\frac{4\Omega}{(2\pi)^{3}}\int\int_{0}^{\infty}dtd^{3}k\left(\left(G_{0}(k)-k^{2}\tilde{u}(k)\right)^{2}Q^{2}(k,t)-\tilde{u}^{2}(k)A(k,t)\right).\label{eq:E2-1}
\end{eqnarray}
 $Q(k,t)$ and $A(k,t)$ are integrals
\begin{eqnarray}
Q(k,t) & = & \frac{1}{(2\pi)^{3}}\int d^{3}p\left\{ \exp(-\left(k^{2}-2\mathbf{p}\cdot\mathbf{k}\right)t)\Theta(1-p)\Theta(|\mathbf{p}-\mathbf{k}|-1)\right\} ,\label{eq:Q-1}\\
A(k,t) & = & \frac{1}{(2\pi)^{6}}\int\int d^{3}pd^{3}q\left\{ \left(k^{2}-(\mathbf{p}-\mathbf{q})\cdot\mathbf{k}\right)^{2}\right.\nonumber\\
&&\left. \Theta(1-p)\Theta(1-q)\Theta(|\mathbf{p}-\mathbf{k}|-1)\Theta(|\mathbf{q}+\mathbf{k}|-1)e^{-\left(2k^{2}-2(\mathbf{p}-\mathbf{q})\cdot\mathbf{k}\right)t}\right\} \nonumber \\
  & = & \frac{1}{4}\frac{\partial^{2}}{\partial t^{2}}\left(Q^{2}(k,t)\right).\label{eq:A2}
\end{eqnarray}
Detailed calculations dealing with $Q(k,t)$ and $A(k,t)$
are presented in Appendix. The results are written as
\begin{eqnarray}
\int_{0}^{\infty}Q^{2}(k,t)dt &=& \frac{1-\ln2}{3(2\pi)^{4}}k+X(k), \label{eq_intQ}\\
\int_{0}^{\infty}A(k,t)dt & = & \frac{1}{2}k^{2}\left(\frac{1}{6\pi^{2}}\right)^{2}T_{2}(k),
\label{eq_intA2}
\end{eqnarray}
where $X(k)$ is defined piecewisely for $k<2$
\begin{eqnarray}
X(k) & = & \frac{1}{2^{6}k(2\pi)^{4}}\frac{1}{5!}\left\{ -704k^{2}-48k^{4}-2048\ln2\right.\nonumber \\
 &  & +\left[1024+960k-160k^{3}+12k^{5}\right]\ln(2+k) \nonumber \\
 &  & \left.+\left[1024-960k+160k^{3}-12k^{5}\right]\ln(2-k)\right\} ,\qquad\mbox{for }k<2,
\end{eqnarray}
and for $k\ge2$
\begin{eqnarray}
X(k) & = & \frac{1}{2^{6}k(2\pi)^{4}}\frac{1}{5!}\left\{ 128k^{3}+2816k
+\left(2k\right)^{3}\left(160-8k^{2}\right)\ln(2k)\right. \nonumber \\
 &  & +\left(2k-4\right)^{3}\left(\left(2k-4\right)^{2}+40k\right)\ln(2k-4) \nonumber \\
 &  & \left. +\left(2k+4\right)^{3}\left(\left(2k+4\right)^{2}-40k\right)\ln(2k+4) \right\} -\frac{1-\ln2}{3(2\pi)^{4}}k,\qquad\mbox{for }k\ge2.
\end{eqnarray}
By using the expression (\ref{eq_intQ}) and (\ref{eq_intA2}) in equation (\ref{eq:E2-1}) we have
\begin{eqnarray}
\frac{E_{21}^{c}}{N} & = & -\frac{4}{\rho'(2\pi)^{3}}\int\int_{0}^{\infty}dtd^{3}k\left(\left(G_{0}(k)-k^{2}\tilde{u}(k)\right)^{2}Q^{2}(k,t)-\tilde{u}^{2}(k)A(k,t)\right)\nonumber\\
 & = & -\frac{3}{2\pi}\int d^{3}k\left\{ \left(G_{0}(\mathbf{k})-k^{2}\tilde{u}(k)\right)^{2}\left(\frac{1-\ln2}{3(2\pi)^{4}}k+X(k)\right)\right.
 \nonumber \\
 &  & \left.-\frac{1}{2}k^{2}\left(\frac{1}{6\pi^{2}}\right)^{2}T_{2}(k)\tilde{u}^{2}(k)\right\} , \label{eq_E21_0}
\end{eqnarray}
where we find that the last term equals $-E_{11}^{c}/N.$ Therefore
we have
\begin{equation}
\frac{E_{21}^{c}}{N}  =  -\frac{1-\ln2}{(2\pi)^{5}}\int d^{3}k\left\{ \frac{k^{5}}{T_{2}^{2}(k)}\tilde{u}^{2}(k)\right\} -\frac{3}{2\pi}\int d^{3}k\left\{ \frac{k^{4}}{T_{2}^{2}(k)}\tilde{u}^{2}(k)X(k)\right\} -\frac{E_{11}^{c}}{N},
\end{equation}
where we have used a simple relation 
\begin{equation}
    G_0(k)-k^2\tilde{u}(k)=-\frac{k^2\tilde{u}(k)}{T_2(k)},
\end{equation}
which can be obtained from equation (\ref{eq_G0}) and (\ref{eq_quadratic}).
As far as the singularity at $k\sim0$ is concerned, the first integrand
is similar to the integrand of $E_{11}^{c}$
\begin{equation}
\frac{k^{5}}{T_{2}^{2}(k)}\tilde{u}^{2}(k)=\left(\frac{4}{3}\right)^{3}T_{2}(k)k^{2}\tilde{u}^{2}(k)+\tilde{u}^{2}(k)O(k^{4}).
\end{equation}
 We can reformulate the expression of $E_{21}^{c}$as
\begin{eqnarray*}
\frac{E_{21}^{c}}{N} & = & -\frac{1-\ln2}{(2\pi)^{5}}\left(\frac{4}{3}\right)^{3}\int d^{3}k\left\{ k^{2}T_{2}(k)\tilde{u}^{2}(k)\right\} \\
 &  & -\frac{1-\ln2}{(2\pi)^{5}}\int d^{3}k\left\{ \frac{k^{5}}{T_{2}^{2}(k)}\tilde{u}^{2}(k)-\left(\frac{4}{3}\right)^{3}T_{2}(k)k^{2}\tilde{u}^{2}(k)\right\} \\
 &  & -\frac{3}{2\pi}\int d^{3}k\left\{ \frac{k^{4}}{T_{2}^{2}(k)}\tilde{u}^{2}(k)X(k)\right\} -\frac{E_{11}^{c}}{N}\\
 & = & \frac{32}{9}(1-\ln2)\frac{E_{11}^{c}}{N}-\frac{E_{11}^{c}}{N}\\
 &  & -\frac{1-\ln2}{(2\pi)^{5}}\int d^{3}k\left\{ \frac{k^{5}}{T_{2}^{2}(k)}\tilde{u}^{2}(k)-\left(\frac{4}{3}\right)^{3}T_{2}(k)k^{2}\tilde{u}^{2}(k)\right\} \\
 &  & -\frac{3}{2\pi}\int d^{3}k\left\{ \frac{k^{4}}{T_{2}^{2}(k)}\tilde{u}^{2}(k)X(k)\right\} ,
\end{eqnarray*}
where the remaining integrals are not any more singular at $k\sim0$
in the high density limit. Then we can use the simple expression (\ref{eq_u_simple})
and the integrals can be numerically evaluated as
\begin{equation}
\frac{E_{21}^{c}}{N}=\frac{32}{9}(1-\ln2)\frac{E_{11}^{c}}{N}-\frac{E_{11}^{c}}{N}-0.00273\alpha^{2}r_{s}^{2},
\end{equation}
and the sum with the first order energy gives 
\begin{equation}
\frac{E_{1}^{c}+E_{21}^{c}}{N}=\alpha^{2}r_{s}^{2}\left[\frac{1-\ln2}{\pi^{2}}\ln(r_{s})-0.0645\right].
\end{equation}

The exchange contraction term of $E_2$  can be written as
\begin{eqnarray}
E_{22} & = & 2\frac{1}{4\Omega^{2}}\sum_{\sigma}\sum_{kk'p}\frac{G(\mathbf{p}-\mathbf{k},\mathbf{p}-\mathbf{k'},\mathbf{k}')G(\mathbf{p},\mathbf{p}-\mathbf{k}-\mathbf{k}',\mathbf{k})}{\frac{1}{2}\left((\mathbf{p}-\mathbf{k})^{2}+(\mathbf{p}-\mathbf{k}')^{2}-p^{2}-(\mathbf{p}-\mathbf{k}-\mathbf{k}')^{2}\right)}\nonumber \\
&&\qquad \Theta(1-p)\Theta(1-|\mathbf{p}-\mathbf{k}-\mathbf{k}'|)\Theta(|\mathbf{p}-\mathbf{k}|-1)\Theta(|\mathbf{p}-\mathbf{k}'|-1)\nonumber \\
 & = & \frac{N}{2\Omega^{2}}\sum_{kk'}\frac{\left(G_{0}(k)-(\mathbf{k}+\mathbf{k}')\cdot\mathbf{k}\tilde{u}(k)\right)\left(G_{0}(k')-(\mathbf{k}'-\mathbf{k})\cdot\mathbf{k}'\tilde{u}(k')\right)}{-\mathbf{k}\cdot\mathbf{k}'}T_{4}(\mathbf{k},\mathbf{k}'),
\end{eqnarray}
where 
\begin{equation}
T_{4}(\mathbf{k},\mathbf{k}')\equiv\frac{2}{N}\sum_{p}\Theta(1-p)\Theta(1-|\mathbf{p}-\mathbf{k}-\mathbf{k}'|)\Theta(|\mathbf{p}-\mathbf{k}|-1)\Theta(|\mathbf{p}-\mathbf{k}'|-1),
\end{equation}
which can be calculated analytically. \cite{luo12}
Therefore in the thermodynamic limit we have
\begin{eqnarray}
\frac{E_{22}}{N} & = & -\frac{1}{2(2\pi)^{6}}\int d^{3}kd^{3}k'\frac{\left(G_{0}(k)-(\mathbf{k}+\mathbf{k}')\cdot\mathbf{k}\tilde{u}(k)\right)\left(G_{0}(k')-(\mathbf{k}'-\mathbf{k})\cdot\mathbf{k}'\tilde{u}(k')\right)}{\mathbf{k}\cdot\mathbf{k}'}T_{4}(\mathbf{k},\mathbf{k}') \nonumber \\
 & = & -\frac{1}{2(2\pi)^{6}}\int_{0}^{\infty}dk\int_{0}^{\infty}dk'\int_{0}^{\pi}d\theta\left\{ 8\pi^{2}k^{2}k'^{2}\sin\theta T_{4}(k,k',\theta) \right.\nonumber\\
 && \left. \frac{\left(G_{0}(k)-(k^{2}+kk'\cos\theta)\tilde{u}(k)\right)\left(G_{0}(k')-(k'^{2}-kk'\cos\theta)\tilde{u}(k')\right)}{kk'\cos\theta}\right\} 
\end{eqnarray}
which can be calculated numerically as
\begin{equation}
   \frac{E_{22}}{N} = 0.01375\alpha^2 r_s ^2.
\end{equation}
 The total correlation energy is finally estimated as
\begin{equation}
\frac{E^{c}}{N}=\alpha^{2}r_{s}^{2}\left[\frac{1-\ln2}{\pi^{2}}\ln(r_{s})-0.05075\right]+o(r_{s}^{2}).
\end{equation}
This result has precisely the same logarithmic term as that of the RPA result, while the constant term is only roughly $7\%$ larger. 

\section{Calculations with the Coulomb-Yukawa factor}\label{CY}
The above calculations are based on the optimised Jastrow factor  for the given system, which is somewhat complicated for the analytic calculations. Can we use some simpler factors instead, which still fulfil the asymptotic conditions?

In quantum Monte Carlo, a Coulomb-Yukawa type function
\begin{equation}
    u(r)=-\frac{\alpha r_s}{f^2 r} (1-e^{-fr}), \qquad f^2=\sqrt{4\pi\rho\alpha r_s},
\end{equation}
is broadly used as the Jastrow factor. Here the expression is reformulated for the re-scaled Hamiltonian (\ref{eq_rescale_2}). The Fourier transformation of $u(r)$ has the following simple form
\begin{equation}
    \tilde{u}(k)=-\frac{4\pi\alpha r_s}{f^2} \left(\frac{1}{k^2}-\frac{1}{k^2 +f^2}\right)
    =-\frac{4\pi \alpha r_s}{k^2(k^2+f^2)},
\end{equation}
where the parameters are determined to fulfil the asymptotic conditions. This factor is much simpler for the integral calculations and serves as a good candidate for the test.

In order to get an idea about the performance of this factor in the second order transcorrelated perturbation calculations, we focus only on the leading terms of the perturbation energies.
The leading term in the first order energy is given by $E^c_{11}$ in equation (\ref{eq_E11_0})
\begin{eqnarray}
 \frac {E_{11}^c} N&=&-\frac{\rho }{2(2\pi)^{3}}\int T_{2}(k)\mathbf{k}^{2}\left(\frac{4\pi \alpha r_s}{k^2(k^2+f^2)}\right)^2d^{3}k\nonumber\\
 &\approx& -\frac{1 }{6\pi^2 (2\pi)^{3}}\int_0^2 \frac 3 4 k^3 \left(\frac{4\pi \alpha r_s}{k^2(k^2+f^2)}\right)^2  4\pi k^2 dk\nonumber\\
 &=& - \frac{\alpha^2 r_s^2}{2\sqrt{3}\pi ^2\alpha^2} \frac{1}{\sqrt{r_s}}+\cdots, 
\end{eqnarray}
which does not lead to a logarithmic term, but rather a more singular $r^{-\frac 1 2}$ term. 

According to equation (\ref{eq_E21_0}), the $E_{11}^c$ term will be completely canceled by the second order energy.  The leading term of the total second order perturbation energy can be evaluated by
\begin{eqnarray}
\frac{E_{11}^c+E_{21}^c} {N} 
 & \approx & -\frac{3}{2\pi}\int d^{3}k\left\{ \left(G_{0}(\mathbf{k})-k^{2}\tilde{u}(k)\right)^{2}\frac{1-\ln2}{3(2\pi)^{4}}k\right\}
 \nonumber \\
 &=& -\frac{3}{2\pi}\int d^{3}k\left\{ \left(\frac{4\pi \alpha r_s}{k^2+f^2}+\frac{4\pi \alpha r_s f^2}{(k^2+f^2)^2}\right)^{2}\frac{1-\ln2}{3(2\pi)^{4}}k\right\}
 \nonumber \\
 &\approx& \alpha^2 r_s^2 \frac{1-\ln 2}{2\pi} \ln r_s,
\end{eqnarray}
which reproduces only half of the exact logarithmic term.

This example shows that the perturbation results relies strongly on the selection of the correlation factor. The current study suggests that a correlation factor optimised for a single determinant reference function (e.g., the Hartree-Fock reference function) serves as a good candidate. This optimisation can be realized by solving either a variational equation, or a transcorrelated equation for the factor. \cite{luo12}

\section{Conclusions and remarks}\label{conclud}
By using a proper Jastrow factor, the leading singularity of the Coulomb potential can be removed from the transcorrelated Hamiltonian. This cures the divergence problem of the MBPT method for metallic systems at low orders. Test calculations on uniform electron gases shows that already at the second order perturbation level, one can get meaningful results for the electron correlations.  We may expect that this method can be developed into an efficient method for metals, curing problems both at the short-range and the long-range. 

The selection of the Jastrow factor is crucial for the calculations. The necessary condition that the factor should fulfil the long- and short-range asymptotic conditions does not guarantee performance. For example, the Coulomb-Yukawa type factor \cite{foulkes01}, which is broadly used in quantum Monte Carlo, does not lead to a meaningful perturbation result.
We also note that this method only cures the divergence at low orders; starting at 4th order, there are still divergent integrals in the perturbation energies.

\clearpage

\appendix

\section{Calculation of $Q(k,t)$}

The function $Q(k,t)$ can be calculated separately for $k\ge 2$ and $k<2$.

\subsection{Case $k\ge2$}

For $k\ge2$, the integration volume is a simple unit sphere
\begin{eqnarray}
Q(k,t) & = & \frac{e^{-k^{2}t}}{(2\pi)^{2}}\int_{0}^{1}\int_{0}^{\pi}\left\{ \exp(2pk\cos\theta\ t)p^{2}\sin\theta\right\} d\theta dp \nonumber\\
 & = & \frac{e^{-k^{2}t}}{(2\pi)^{2}(2kt)^{3}}\left\{ 2kt\left(e^{2kt}+e^{-2kt}\right)-\left(e^{2kt}-e^{-2kt}\right)\right\} . \label{eq_q_1}
\end{eqnarray}
In the calculation of $\int_{0}^{\infty}A(k,t)dt$
integrals we also need to evaluate the leading orders of $Q$
\begin{eqnarray*}
Q(k,t) & = & \frac{e^{-k^{2}t}}{(2\pi)^{2}(2kt)^{3}}\left\{ 4kt\left(1+2k^{2}t^{2}+O(t^{4})\right)-\left(4kt+(2kt)^{3}/3+O(t^{5})\right)\right\} \\
 & = & \frac{e^{-k^{2}t}}{6\pi^{2}}\left\{ 1+O(t^{2})\right\} ,
\end{eqnarray*}
and thus
\begin{eqnarray}
Q(k,0) & = & \frac{1}{6\pi^{2}},\\
\frac{\partial}{\partial t}\left(Q^{2}(k,t)\right)_{t=0} & = & -2k^{2}\left(\frac{1}{6\pi^{2}}\right)^{2}. 
\label{eq_d2q_1}
\end{eqnarray}

\subsection{case $k<2$}
For $k<2$, the integration volume is the unit sphere centered at $\mathbf{0}$
subtracts its intersection with the unit sphere centered at $\mathbf{k}$
\begin{eqnarray}
Q(k,t) & = & \frac{e^{-k^{2}t}}{(2\pi)^{2}(2kt)^{3}}\left\{ 2kt\left(e^{2kt}+e^{-2kt}\right)-\left(e^{2kt}-e^{-2kt}\right)\right\} \nonumber\\
 &  & -\frac{1}{(2\pi)^{2}}\int_{0}^{\sqrt{1-k^{2}/4}}\int_{-\sqrt{1-r^{2}}+k/2}^{\sqrt{1-r^{2}}-k/2}dzdr\left\{ r\exp(2kzt)\right\} \nonumber \\
 & = & \frac{1}{(2\pi)^{2}(2kt)^{3}}\left\{ (2kt+1)\left(e^{-2kt-k^{2}t}-e^{-2kt+k^{2}t}\right)+2k^{2}t\right\} \label{eq_q_2}.
\end{eqnarray}
Again we need the leading terms of the Taylor expansion of $ Q(k,t) $ with respect to $t$, which can be calculated as
\begin{equation}
 Q(k,t)= \frac{1}{6\pi^{2}}\left\{ T_{2}(k)-k^{2}t+O(t^{2})\right\} .
\end{equation}
Then we can get 
\begin{eqnarray}
Q(k,0) & = & \frac{1}{6\pi^{2}}T_{2}(k),\\
\frac{\partial}{\partial t}\left(Q^{2}(k,t)\right)_{t=0} & = & -2k^{2}\left(\frac{1}{6\pi^{2}}\right)^{2}T_{2}(k).
\label{eq_d2q_2}
\end{eqnarray}
Here we find equation (\ref{eq_d2q_1}) and equation (\ref{eq_d2q_2}) are identical, since for $k>2$, $T_2(k)=1.$

\section{Calculation of $\int_{0}^{\infty}Q^{2}(k,t)dt$ and $\int_{0}^{\infty}A(k,t)dt $}
\subsection{case $k\ge 2$}
According to equation (\ref{eq_q_1})
\begin{eqnarray*}
\int_{0}^{\infty}Q^{2}(k,t)dt & = & \frac{1}{2^{6}(2\pi)^{4}}\int_{0}^{\infty}dt\frac{e^{-2k^{2}t}}{(kt)^{6}}\left(2kt\left(e^{2kt}+e^{-2kt}\right)-\left(e^{2kt}-e^{-2kt}\right)\right)^{2}\\
 & = & \frac{1}{2^{6}k(2\pi)^{4}}\int_{0}^{\infty}dx\frac{e^{-2kx}}{x^{6}}\left(2x\left(e^{2x}+e^{-2x}\right)-\left(e^{2x}-e^{-2x}\right)\right)^{2}.
\end{eqnarray*}
This integral has the form $\int_{0}^{\infty}\frac{F(x)}{x^{6}}dx$.
Since $Q(k,t)$ is finite at $t=0,$ the leading term of $F(x)$
has the order $x^{6}$. We can then use the simple relation
\begin{equation}
    \int_{0}^{\infty}\frac{F(x)}{x^{6}}dx=\int_{0}^{\infty}\frac{F^{(5)}(x)}{5!x}dx.
\end{equation}
The calculation with $F^{(5)}$ is lengthy but straightforward, and in the end we are left with only the following type integrals
\begin{equation*}
\int_{0}^{\infty}xe^{-ax}dx=a^{-2},\quad\int_{0}^{\infty}e^{-ax}dx=a^{-1},\quad\int_{0}^{\infty}\frac{e^{-ax}-e^{-bx}}{x}dx=\ln\frac{b}{a}.
\end{equation*}
The final result of the integral is
\begin{eqnarray*}
\int_{0}^{\infty}Q^{2}(k,t)dt 
 & = & \frac{1}{2^{6}k(2\pi)^{4}}\frac{1}{5!}\left\{ 128k^{3}+2816k
 +\left(2k\right)^{3}\left(160-8k^{2}\right)\ln(2k)\right.\\
 &  &+\left(2k-4\right)^{3}\left(\left(2k-4\right)^{2}+40k\right)\ln(2k-4)\\
 & & \left.+\left(2k+4\right)^{3}\left(\left(2k+4\right)^{2}-40k\right)\ln(2k+4)\right\}
\end{eqnarray*}

The integral $\int_{0}^{\infty}A(k,t)dt$ can be simply calculated by partial integration
\begin{eqnarray*}
\int_{0}^{\infty}A(k,t)dt & = & \int_{0}^{\infty}dt\left\{ \frac{1}{4}\frac{\partial^{2}}{\partial t^{2}}\left(Q^{2}(k,t)\right)\right\} \\
 & = & -\frac{1}{4}\frac{\partial}{\partial t}\left(Q^{2}(k,t)\right)_{t=0}\\
 & = & \frac{k^{2}}{2}\left(\frac{1}{6\pi^{2}}\right)^{2}.
\end{eqnarray*}

\subsection{case $k< 2$}
In the same way as above, the integral $\int_{0}^{\infty}Q^{2}(k,t)dt$ can be calculated as
\begin{eqnarray*}
\int Q^{2}(k,t)dt
 & = & \frac{1}{(2\pi)^{4}}\int_{0}^{\infty}dt\frac{1}{(2kt)^{6}}\left\{ (2kt+1)\left(e^{-2kt-k^{2}t}-e^{-2kt+k^{2}t}\right)+2k^{2}t\right\}^2\\
 & = & \frac{1}{2^{6}k(2\pi)^{4}}\frac{1}{5!}\left\{ 1856k^{2}-48k^{4}-\left[2048+2560k^{2}\right]\ln2 \right.\\
 &  & +\left[1024+960k-160k^{3}+12k^{5}\right]\ln(2+k)\\
 &  & \left.+\left[1024-960k+160k^{3}-12k^{5}\right]\ln(2-k)\right\}.
\end{eqnarray*}
For the calculation of the logarithmic term in $E_{2}^{c}$, we split
out the leading order term of the above integral and write it as
\begin{equation*}
\int_{0}^{\infty}Q^{2}(k,t)dt=\frac{1-\ln2}{3(2\pi)^{4}}k+X(k),
\end{equation*}
with $X(k)\sim O(k^{2}) $ being the high order part
\begin{eqnarray*}
X(k) & = & \frac{1}{2^{6}k(2\pi)^{4}}\frac{1}{5!}\left\{ -704k^{2}-48k^{4}-2048\ln2\right.\\
 &  & +\left[1024+960k-160k^{3}+12k^{5}\right]\ln(2+k)\\
 &  & \left.+\left[1024-960k+160k^{3}-12k^{5}\right]\ln(2-k)\right\} .
\end{eqnarray*}
The integral on $A$ can also be calculated simply by
partial integration
\begin{eqnarray*}
\int_{0}^{\infty}A(k,t)dt & = & \int_{0}^{\infty}dt\left\{ \frac{1}{4}\frac{\partial^{2}}{\partial t^{2}}\left(Q^{2}(k,t)\right)\right\} \\
 & = & -\frac{1}{4}\frac{\partial}{\partial t}\left(Q^{2}(k,t)\right)_{t=0}\\
 & = & \frac {k^2} 2 \left(\frac{1}{6\pi^{2}}\right)^{2}T_{2}(k).
\end{eqnarray*}

\bibliography{tcpt}
\end{document}